\documentclass[final,5p,times,twocolumn]{elsart4}

\usepackage{natbib}

\usepackage{amssymb}
\usepackage{bm}
\usepackage{graphicx}% Include figure files

\begin{document}

\begin{frontmatter}

% Title, authors and addresses

% use the thanksref command within \title, \author or \address for footnotes;
% use the corauthref command within \author for corresponding author footnotes;
% use the ead command for the email address,
% and the form \ead[url] for the home page:
% \title{Title\thanksref{label1}}
% \thanks[label1]{}
% \author{Name\corauthref{cor1}\thanksref{label2}}
% \ead{email address}
% \ead[url]{home page}
% \thanks[label2]{}
% \corauth[cor1]{}
% \address{Address\thanksref{label3}}
% \thanks[label3]{}

\title{A hierarchical research by \\
large-scale and {\it ab initio} electronic structure theories \\
-- Si and Ge cleavage and stepped (111)-2$\times$1 surfaces -- 
}

% use optional labels to link authors explicitly to addresses:
% \author[label1,label2]{}
% \address[label1]{}
% \address[label2]{}

\author{T. Hoshi\corauthref{cor1}\thanksref{label1}\thanksref{label2}}
 \ead{hoshi@damp.tottori-u.ac.jp}
\author{M. Tanikawa \thanksref{label1}}
\author{A. Ishii \thanksref{label1}\thanksref{label2}}

\corauth[cor1]{corresponding author \\
\quad Takeo Hoshi \\
\quad Tottori University \\
\quad 4-101 Koyama-Minami, Tottori 680-8552 Japan}

\address[label1]{Department of Applied Mathematics and Physics, Tottori University, Tottori, Japan}
\address[label2]{Core Research for Evolutional Science and Technology , 
Japan Science and Technology Agency (JST-CREST), Saitama, Japan.}

%\corauth[cor2]{Present address: 
%%Canon Inc., Analysis technology center, 
%5-1 Morinosato-Wakamiya, Atsugi-shi, Kanagawa 243-0193, Japan}

\begin{abstract}
The {\it ab initio} calculation with the density functional theory
and plane-wave bases is carried out for
stepped Si(111)-2$\times$1 surfaces
that were predicted in a cleavage simulation
by the large-scale (order-$N$) electronic structure theory 
(T. Hoshi, Y.  Iguchi and T. Fujiwara,  
Phys. Rev. B {\bf 72} (2005) 075323). 
The present {\it ab initio} calculation confirms 
the predicted stepped structure and 
its bias-dependent STM image.
Moreover, two (meta)stable step-edge structures 
are found and compared. 
The investigation is carried out also for Ge(111)-2$\times$1 surfaces,
so as to construct a common understanding among elements.
The present study demonstrates the general importance
of the hierarchical research between
large-scale and {\it ab initio} electronic structure theories.
\end{abstract}


\end{frontmatter}

\pagebreak

% main text

\section{Introduction \label{INTRO}}

A promising theoretical approach to nano materials
is a hierarchical or multiscale research
by large-scale and {\it ab initio} electronic structure calculations,
since the research can handle, seamlessly,
nano systems with a wide range of length scale. 
Nowadays, 
several {\it ab initio} calculation methods 
are well established, typically, for systems with $10^1-10^2$ atoms
and a popular method is the one with
the density functional theory and plane-wave bases (DFT-PW calculation)
\cite{CP}.
Electronic structure methods for larger systems, on the other hand,  
have been proposed by many groups. 
A set of fundamental methodologies of 
large-scale calculation and related theories were developed. 
\cite{HOSHI2000, HOSHI2001, HOSHI2003, GESHI2004,TAKAYAMA2004,HOSHI2005,
TAKAYAMA2006, HOSHI2006-PHYSICA-B, HOSHI2006, HOSHI2007, IGUCHI2007,
SHINAOKA,YAMAMOTO2008,HOSHI-JPCM2009, HOSHI-JPCM2009-HELICAL,HOSHI-HNP,HOSHI-JPCS-NPD}
Among them,
the large-scale calculations were realized 
by \lq order-$N$' algorithms and/or parallel computations.
Samples with $10^2$-$10^7$ atoms were calculated 
with tight-binding form Hamiltonians. 
Now the program code has the name of 
\lq ELSES' (=Extra-Large-Scale Electronic Structure calculation;
http://www.elses.jp/) .
The code was used particularly
in the researches   
of silicon cleavage (explained hereafter), 
helical multishell gold nanowires 
\cite{IGUCHI2007, HOSHI-JPCM2009-HELICAL,HOSHI-HNP} 
(See Ref.~\cite{KONDO2000} as experiments) and 
nano-polycrystalline diamond \cite{HOSHI-JPCS-NPD}
(See Ref. ~\cite{IRIFUNE-NPD-NATURE2003} as experiments).
Other methodologies for large-scale calculations can be found,
for example, in the literatures of Refs. \cite{HOSHI2006,HOSHI-JPCM2009}.

The present paper focuses on 
the cleavage dynamics and resultant 
stepped Si(111)-2x1 surfaces,  
as a hierarchal research by the large-scale 
calculations and the DFT-PW calculations.
This paper organized as follows;
Section \ref{SEC-CLEAVAGE} is devoted to a review 
of the Si(111)-2x1 surface 
and the result of the large-scale calculations in previous papers.
In Sec.~\ref{ABINITIO}, 
the results of DFT-PW calculations are presented 
for stepped Si and Ge(111)-2x1 surfaces,
so as to construct a hierarchical research.
Finally, a summary is given.

\section{Si(111)-2$\times$1 cleavage mode and its large-scale calculations}
\label{SEC-CLEAVAGE}

The Si(111)-2x1 surface, called Pandey structure, 
\cite{PANDEY}
appears on cleaved samples and 
is of great importance
in fracture science and surface science. 
See papers 
\cite{FEENSTRA87, PARRINELLO,
FEENSTRA86, SPENCE93, ROHLFING1999, GUMBSCH, 
HOSHI2005,KERMODE2008, INAMI2007} 
and references therein.
In the Pandey structure, 
threefold-coordinated surface atoms 
form a zigzag chain, like a part of graphen sheet.
The zigzag chains on the surface are tilted and 
a pioneering DFT-PW calculation shows
the transformation process between the two tilted structures.
\cite{PARRINELLO}
Cleaved Ge samples also contain the (111)-2$\times$1 surface.
See the above papers and reference therein.

%-%-%-%-%-%-%-%-%-%-%-%-%-%-%-%-%-%-%-%-%-%-%-%-%-%-%-%-%-%-%
\begin{figure}[tbh]
\begin{center}
   \includegraphics[width=7cm]{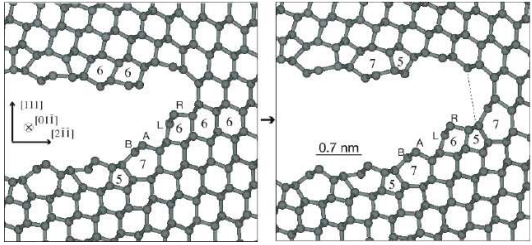}
\end{center}
\caption{
Close up of the snapshots the cleavage dynamics with a 10-nm-scale system,
by the large-scale electronic structure calculation. 
\cite{HOSHI2005}
A step is formed on the (111)-2$\times$1 surface.
The time interval between the left and right snapshots is 0.6 ps. 
See the text for details.
Several five-, six-, seven-membered rings are marked
as \lq 5', \lq 6' and \lq 7', respectively. 
The dashed lines indicate the initial (crystalline) bonds. 
}
\label{FIG-FRAC}
\end{figure}%
%-%-%-%-%-%-%-%-%-%-%-%-%-%-%-%-%-%-%-%-%-%-%-%-%-%-%-%-%-%-%

%-%-%-%-%-%-%-%-%-%-%-%-%-%-%-%-%-%-%-%-%-%-%-%-%-%-%-%-%-%-%
\begin{figure}[tbh]
\begin{center}
  \includegraphics[width=7cm]{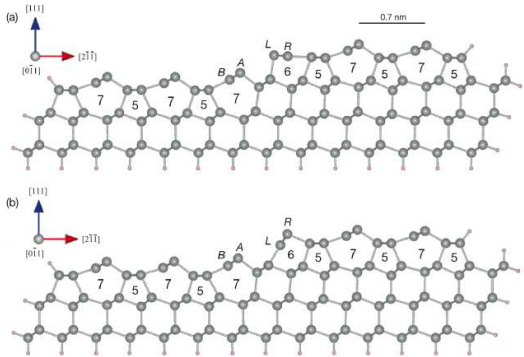}
  %\vspace{7cm}
\end{center}
\caption{
Stepped Si(111)-2$\times$1 structures 
calculated by the DFT-PW calculation. 
The small balls are hydrogen terminations. 
(a) \lq $L$-type' and (b) \lq $R$-type' step-edge structures
are obtained as stable structures.
See the text for details.
}
\label{FIG-STEPPED-VASP}
\end{figure}%
%-%-%-%-%-%-%-%-%-%-%-%-%-%-%-%-%-%-%-%-%-%-%-%-%-%-%-%-%-%-%

Large-scale order-$N$ calculations were carried out
for  the cleavage dynamics 
and the local electronic density of states (LDOS) 
on the resultant cleaved surface. 
\cite{HOSHI2003, HOSHI2005, HOSHI2006-PHYSICA-B, HOSHI2006, HOSHI2007}
The cleavage dynamics was carried out
by  the generalized Wannier-state method 
\cite{HOSHI2000, HOSHI2001,HOSHI2003, GESHI2004, 
HOSHI2005, HOSHI2006-PHYSICA-B, HOSHI2006, HOSHI2007} and
the LDOS calculation was realized
by a Krylov-subspace algorithm for the Green's function.
\cite{TAKAYAMA2004, HOSHI2006-PHYSICA-B,TAKAYAMA2006}

Figure ~\ref{FIG-FRAC} shows an example of 
the cleavage dynamics with a 10-nm-scale sample
or $10^4$ atoms.\cite{HOSHI2005}
The  (111)-2$\times$1 reconstruction 
appears on the cleaved surface.
Several results contain step formations on 
the  (111)-2 $\times$ 1 surface, as shown in Fig.~\ref{FIG-FRAC}.
In Fig.~\ref{FIG-FRAC}, 
several five-, six-, seven-membered rings are marked
as \lq 5', \lq 6' and \lq 7', respectively, as a guide for eye. 
The 2$\times$1 reconstruction appears,
when the two successive six-membered rings
are transformed into a set of five- and seven-membered rings.
A characteristic six-membered ring appears at the step-edge region.

Surface atoms on the flat region,
form tilted zigzag chains of threefold-coordinated atoms,
as explained in the first paragraph of the present section.
For example,
the $A$ atom in Fig.~\ref{FIG-FRAC} is placed on the vacuum-side (upper) region
and the $B$ atom is placed on the bulk-side (lower) region.
The zigzag chain with the reverse tilting is also obtained. 
It is remarkable that 
the surface atoms at the step-edge region,
marked as \lq $L$' and \lq R',
also form a  zigzag chains of threefold-coordinated atoms, 
as on the flat region.
One can consider, therefore, that
the two tilted structures may be stable 
at the step-edge region, as on the flat region,
although the simulation results 
contain only one tilted structure in which 
the $R$ atoms is placed on the vacuum-side region and 
the $L$ atoms is on the bulk-side region.
Moreover, 
the calculated LDOS indicates
a bias-dependent image of
the scanning tunnelling spectroscopy (STM) 
both on the flat region and the step-edge regions.
The peaks in the LDOS profile 
are located at almost the same positions 
in the step-edge atoms, such as the $L$ and $R$ atoms, 
and the atoms on the flat region, such as the $A$ and $B$ atoms.
A bias-dependent STM image was observed experimentally
for the flat region.
\cite{FEENSTRA86}
The calculation predicts the same bias-dependent image
on the step-edge region.
The same step-edge structures and its bias-dependent STM image 
were suggested also for Ge cases,
because of the similarity of electronic structure.
\cite{HOSHI2005}

%%%%%%%%%%%%%%%%%%%%%%%%%%%%%%%%%%%%
\section{{\it Ab initio} calculations \label{ABINITIO}}

The properties of the stepped Si(111)-2$\times$1 surface,
predicted by the large-scale calculation, 
were confirmed by the DFT-PW calculation with a smaller sample.
The Vienna {\it ab initio} simulation package (VASP)  
was used with the projector augmented wave method 
and the local density approximation. \cite{VASP}

Figure \ref{FIG-STEPPED-VASP} shows
the calculated sample  
that contains 96 silicon atoms in the simulation cell.  
The sample corresponds to a \lq close up' 
of the step-edge region of the large-scale calculation.
The sample is periodic only in the $[0\bar{1}1]$ direction.
and the sample boundary 
is terminated by artificial hydrogen atoms.

Structures are optimized with several initial guesses and
the two tilted step-edge structures in Fig.~\ref{FIG-STEPPED-VASP}
are obtained as stable ones. 
The site names of $A$, $B$, $L$ and $R$ in  Fig.~\ref{FIG-STEPPED-VASP}
are consistent to those in Fig.~\ref{FIG-FRAC}.
Hereafter, the structure in  Fig.~\ref{FIG-STEPPED-VASP} (a) is called 
\lq L-type' step-edge structure,
since the $L$ atom, the left atom of the step edge region,
is located at the vacuum-side region
and the $R$ atom, the right atoms, is located at the bulk-side region.
The structure in  Fig.~\ref{FIG-STEPPED-VASP} (b)
indicates the reverse tilting at the step-edge region
and is called \lq R-type' structure.
The energy of the $L$-type structure is lower 
than that of the $R$-type structure
by $\Delta E \equiv E_{\rm R}  - E_{\rm L} = 0.19$ eV 
per atom pair of the zigzag chain.

Additional calculations were carried out, 
so as to clarify physical discussions for the two step-edge structures.
First, the structures were calculated 
with different tilting structures on the flat region,
in which the zigzag chains on the flat region,
such as the $A$ and $B$ atoms in Fig.~\ref{FIG-STEPPED-VASP}, 
form the reverse tilting structure. 
The calculation result indicates that 
the above energy difference 
  ($\Delta E = 0.19$ eV) is independent 
on the tilting strctures of the flat region
and that the above energy difference stems from 
the local tilting structure of the step-edge region. 
Second, 
the energy difference
was calculated also by the tight-binding form Hamiltonian 
used in the large-scale calculation 
and gives $\Delta E = 0.22$eV,
which agrees satisfactorily with the above {\it ab initio} calculation.

Although the existence of the two stable structures at the step-edge regions
is reasonable, as discussed in the previous section, 
it is interesting that the cleavage simulation of Fig.~\ref{FIG-FRAC}
contains the \lq R-type' structure,
the higher energy structure of the two. 
The appearance of the higher energy structure in the cleavage simulation
can be understood,
since the $R$ atom is located on an upper position than the $L$ atom,
in an ideal diamond structure.
The simulation realizes the dynamics with the time scale in 10 ps and 
it might be difficult to reproduce
the transformation process 
from the \lq R-type' structure into the \lq L-type' one.

%-%-%-%-%-%-%-%-%-%-%-%-%-%-%-%-%-%-%-%-%-%-%-%-%-%-%-%-%-%-%
\begin{figure}[thb]
\begin{center}
  \includegraphics[width=7cm]{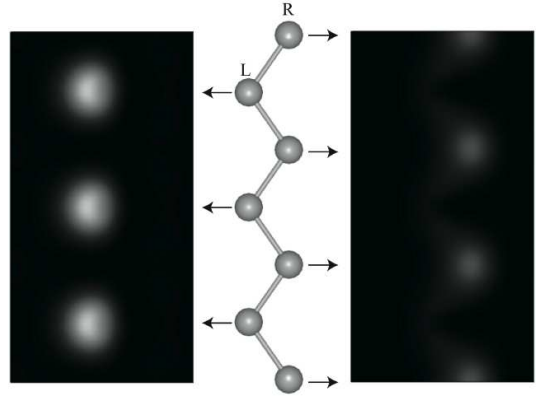}
\end{center}
\caption{
The bias-dependent STM images at the step-edge atoms 
in the \lq L-type' structure.
The image was
calculated with the DFT-PW calculation and
the bias value $\varepsilon_{\rm V}$ is set to be 
$\varepsilon_{\rm V} = \varepsilon_{\rm F} $ -0.7 eV 
and $\varepsilon_{\rm F} $ + 0.7 eV
in the left and right panels, respectively.
The middle panel shows the top view 
of the zigzag chain of 
threefold-coordinated surface atoms, such as the $L$ and $R$ atoms in 
Fig.~\ref{FIG-STEPPED-VASP}(a).
}
\label{FIG-STM-VASP}
\end{figure}
%-%-%-%-%-%-%-%-%-%-%-%-%-%-%-%-%-%-%-%-%-%-%-%-%-%-%-%-%-%-%

In Fig.~\ref{FIG-STM-VASP},
the STM images are calculated for the step-edge region 
of the \lq L-type' structure.
The calculation details are written in the caption.
An occupied surface state is localized 
on the vacuum-side atom (the $L$ atom) 
and an unoccupied surface state is localized 
on the bulk-side atom (the $R$ atom).
The calculated STM images for the flat region
shows the same bias-dependency.
These results are consistent to those of the large-scale calculation.
Here it is noteworthy that the STM image 
for the \lq R-type' structure
should be discussed more carefully,
both in theory and experiment, 
since the height of the $L$ and $R$ atoms 
are different  significantly, by approximately 1 \AA. 
When STM images were calculated as the  constant height calculations,
like those for the  \lq L-type' structure in Fig.~\ref{FIG-STM-VASP},
and the probed height was set to be slightly higher than the $R$ atom,
the calculated images contain only the images for the $R$ atom, like the right panel of 
Fig.~\ref{FIG-STM-VASP}, both for the positive and negative bias values,
since the $R$ atom is placed near the probed height. 

%-%-%-%-%-%-%-%-%-%-%-%-%-%-%-%-%-%-%-%-%-%-%-%-%-%-%-%-%-%-%
\begin{figure}[thb]
\begin{center}
  \includegraphics[width=7cm]{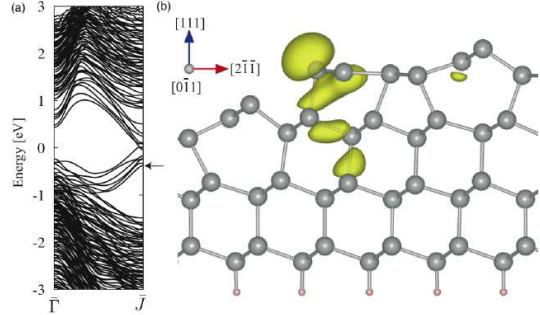}
  %\vspace{7cm}
\end{center}
\caption{
(a) The band structure for
the \lq L-type' step-edge structure shown 
in Fig.~\ref{FIG-STEPPED-VASP}(a).
(b) The profile of the eigen state 
indicated by the arrow at the $\bar{J}$ point in (a).
}
\label{fig-4-step-edge-wfn}
\end{figure}%
%-%-%-%-%-%-%-%-%-%-%-%-%-%-%-%-%-%-%-%-%-%-%-%-%-%-%-%-%-%-%

Figure ~\ref{fig-4-step-edge-wfn} shows 
(a) the band structure and (b) a characteristic localized wavefunction
for the \lq L-type' step-edge structure.
These results are quite similar to those of the flat Si(111)-2x1 surface,
since the zigzag chain of 
threefold-coordinated atoms appears commonly in the two cases.
The band structure of 
the flat Si(111)-2x1 surface  is found, for example, 
in Ref.~\cite{ROHLFING1999}, in which 
they use the GW approximation,
an advanced electronic structure theory beyond DFT.
The surface states are separated from the bulk bands at the $\bar{J}$ point,
as a common property among
the band structures of 
the flat surface (Fig. 1 of Ref.~\cite{ROHLFING1999})
and the stepped surface (Fig.~\ref{fig-4-step-edge-wfn}(a)).
Figure ~\ref{fig-4-step-edge-wfn}(b)
shows a characteristic wavefunction localized at the step-edge region.
The corresponding eigen level is placed at 
the lowest surface state level the $\bar{J}$ point.
The level is indicated by the arrow in Fig.~\ref{fig-4-step-edge-wfn}(a).
The wavefunction is a $\pi$-like state 
located on the vacuum-side ($L$) atom at the step edge region
and the same property is seen on 
the wavefunction on the flat surface
(Fig.2(a) of Ref.~\cite{ROHLFING1999}).
The energy gap between the occupied and unoccupied levels
vanishes at the $\bar{J}$ point  in Fig.~\ref{fig-4-step-edge-wfn}(a).
It is speculated that the gap is underestimated 
artificially, as usually within DFT calculations, 
and a finite gap should appear in a real system,
as that of Ref.~\cite{ROHLFING1999}

The same investigation was carried out 
also for the stepped Ge(111)-2x1 surfaces.
The energy of the $L$-type step-edge structure is lower 
than that of the $R$-type structure, as in the Si case.
The energy difference between them
is comparable to that in the Si case 
($\Delta E = 0.23$ eV).
Therefore, 
the appearance of the stepped (111)-2x1 structure 
should be understood as a common property 
of the Si and Ge cases,
which is consistent to the discussion in the previous section.

\section{Summary}

A hierarchical research is presented
for the stepped Si and Ge (111)-2$\times$1 surface 
in the stable cleavage mode.
The present DFT-PW calculation validates
the predictions 
in our previous calculation 
with  the large-scale order-$N$ theory \cite{HOSHI2005}
for (i) the stepped surface structure
and (ii) its bias-dependent STM image.

The hierarchical research as a combined study 
of large-scale calculations with large (10-nm-scale) systems 
and finer calculations with smaller systems
is general and is applicable to various nanomaterials, 
since all the theories are founded within a general framework
of quantum mechanics for electron.

\end{document}